\documentclass[a4paper]{article}
\usepackage{graphicx}
\usepackage{amsmath}
\usepackage{amsfonts}
\usepackage{amssymb}
\begin{document}

\title{On fluid/wall slippage}
\author{P.\ G.\ de Gennes\\Coll\`{e}ge de France, 11 place Marcelin\ Berthelot\\75231 Paris Cedex 05, France\\E.mail: pgg@espci.fr}
\maketitle
\begin{abstract}
Certain (non polymeric) fluids show an anomalously low friction when flowing
against well chosen solid walls.\ We discuss here one possible explanation,
postulating that a gaseous film of small thickness $h$ is present between
fluid and wall. When $h$ is smaller than the mean free path $\ell$ of the gas
(Knudsen regime) the Navier length $b$ is expected to be independent of $h$
and very large (microns).
\end{abstract}

\section{Introduction}

The standard boundary condition for fluid flow along a wall is a no slip
condition. If the wall is at rest, the tangential fluid velocity at the wall
vanishes.\ The validity of this postulate was already checked in pioneering
experiments by Coulomb \cite{coulomb}.

\begin{figure}
[h]
\begin{center}
\includegraphics[
height=2.348in,
width=2.3402in
]%
{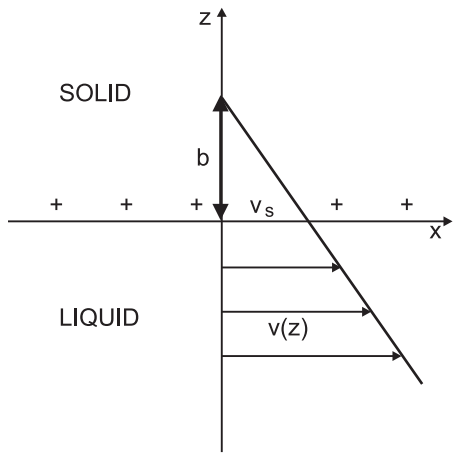}%
\caption{Definition of the Navier length $b$ for simple shear flow near a
solid wall}%
\label{fig1}%
\end{center}
\end{figure}

However, the spatial resolution available to Coulomb was limited. In our days,
we characterise the amount of slip by an extrapolation length $b$ (the Navier
length): the definition of $b$ is explained on fig. \ref{fig1}. The length $b$
can be related to the surface friction coefficient $k$ defined as follows: the
shear stress $\sigma$ induces at the wall a surface velocity v$_{s}$:%

\begin{equation}
\sigma=k\text{v}_{s} \label{eq1}%
\end{equation}%

%TCIMACRO{\FRAME{fhFU}{2.3402in}{2.348in}{0pt}{\Qcb{Definition of the Navier
%length $b$ for simple shear flow near a solid wall}}{\Qlb{fig1}}%
%{fig1.eps}{\special{ language "Scientific Word";  type "GRAPHIC";
%maintain-aspect-ratio TRUE;  display "PICT";  valid_file "F";
%width 2.3402in;  height 2.348in;  depth 0pt;  original-width 4.6241in;
%original-height 4.6389in;  cropleft "0";  croptop "1";  cropright "1";
%cropbottom "0";  filename 'fig1.eps';file-properties "XNPEU";}}}%
%BeginExpansion
%EndExpansion

Equating this to the viscous shear stress in the fluid (of velocity $\eta$),
we get:%

\begin{equation}
\sigma=k\text{v}_{s}=\eta\left|  \frac{d\text{v}(z)}{dz}\right|  \label{eq2}%
\end{equation}

where v$(z)$ is the fluid velocity, increasing linearly with the distance from
the wall $z$. When compared with fig. \ref{fig1}, we see that eq. \ref{eq2} gives:%

\begin{equation}
b=\frac{\eta}{k} \label{eq3}%
\end{equation}

For most practical situations, with simple fluids (made of small molecules
with a diameter $a$), we expect a very small Navier length $b\sim a$: this has
been verified in classical experiments, using a force machine under slow shear
\cite{drake}.

1) A polymer melt, facing a carefully prepared surface, with grafted chains
which are chemically identical to the melt: here, when $\sigma$ becomes larger
than a certain critical value $\sigma^{\ast}$, the Navier length jumps to high
values ($\sim$50 microns) \cite{leger}.\ This can plausibly be explained in
terms of polymer dynamics \cite{brochard}: at $\sigma>\sigma^{\ast}$, the
grafted chains are sufficiently elongated by the flow to disentangle from the
moving chains.

2) With water flowing in thin, hydrophobic, capillaries, there is some early
qualitative evidence for slippage \cite{churaev}, \cite{blake}.

3) The role of the hydrophobic surfaces has been analysed theoretically by
J.\ L.\ Barrat and coworkers -using analysis plus simulations \cite{barrat},
\cite{barrat2}. They show that the first layer of ''water'' molecules is
depleted in the presence of a hydrophobic wall: this can lead in their case to
$b\sim15$ molecular diameters in a typical case, for a contact angle
$\theta=150%
%TCIMACRO{\UNICODE{0xb0}}%
%BeginExpansion
{{}^\circ}%
%EndExpansion
$, corresponding to strong hydrophobicity.

4) Recent experiments by Hervet, L\'{e}ger and coworkers are based on a local
photobleaching technique (using evanescent waves) which probes the velocity
field v$(z)$ within 50 nanometers of the surface \cite{pit}.\ They studied
hexadecane flowing on sapphire: with bare sapphire, no slip was detected.\ On
a fully grafted (methylated) sapphire, the slip length was found to be large
(175 nm) and independent of $\sigma$ in the observed range.\ This is a
surprise: \ \ a) we do not expect any hydrophobic effect here \ \ b) even if
it was present, the $b$ values are much larger than the Barrat-Bocquet estimates.

5) Similar (as yet unpublished) results have been obtained by Zhu and Granick
(using a force machine) and by Tabeling (using etched microcapillaries).

The results \cite{brochard} and \cite{churaev} are unexpected and stimulating.
This led us to think about unusual \ processes which could take place near a
wall.\ In the present note, we discuss one (remote) possibility: the formation
of a gaseous film at the solid/liquid interface.

The source of the film is unclear: when the contact angle is large
($\theta\rightarrow180%
%TCIMACRO{\UNICODE{0xb0}}%
%BeginExpansion
{{}^\circ}%
%EndExpansion
$), a type of flat bubble can form at the surface with a relative low
energy.\ But this energy is still high when compared to $kT.$

We first considered flat vapor bubbles generated at the solid fluid interface
by thermal fluctuations.\ But the typical thickness of these bubbles turns out
to be very small: of order ($kT/\gamma)^{1/2}$ (where $T$ is the temperature,
and $\gamma$ the surface tension), even when $\theta$ is closed to 180$%
%TCIMACRO{\UNICODE{0xb0}}%
%BeginExpansion
{{}^\circ}%
%EndExpansion
$: thus, this brings us back to the Barrat picture, with deplection in one
first layer of the liquid.

Another possible source is an external gas dissolved in the liquid, up to
metastable concentrations: this would then nucleate bubbles preferentially
near the wall if $\theta>90%
%TCIMACRO{\UNICODE{0xb0}}%
%BeginExpansion
{{}^\circ}%
%EndExpansion
.$

In section 2, we simply assume that there exists a uniform gas film of
thickness $h$, larger than the molecule size $a$, but smaller than the mean
free path $\ell$ in the gas.\ We calculate the Navier length $b$ for this
case, and find large values. The possible meaning of this result is discussed
in section 3.

\section{A gas film in the Knudsen regime}

\begin{figure}
[h]
\begin{center}
\includegraphics[
height=1.6985in,
width=2.6109in
]%
{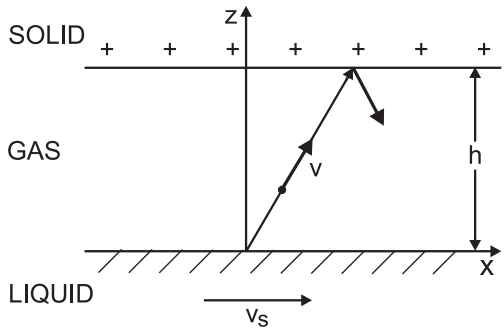}%
\caption{A gas molecule leaves the liquid with the velocity $\overset
{\rightarrow}{\text{v}}$, and transmits (on the average) a momentum $mV$ to
the solid surface.}%
\label{fig2}%
\end{center}
\end{figure}

The situation is shown on fig. \ref{fig2}.\ In the Knudsen regime, gas
molecules travel freely in the film and hit the boundaries. We assume that
there is no specular reflection on either boundary. Then, a molecule leaving
the liquid will have a gausian velocity distribution for the tangential
component v$_{x}$, with the peak of this gausian at velocity v$_{s}$ (the
translational velocity at the liquid surface).\ On the average, it will
transmit a momentum $m$v$_{s}$ to the solid. The number of such hits per
second is $\rho/m\,$v$_{z}$, where $\rho$ is the gas density, $m$ the
molecular mass, and v$_{z}$ the normal component of velocity.%

%TCIMACRO{\FRAME{fhFU}{2.6109in}{1.6985in}{0pt}{\Qcb{A gas molecule leaves the
%liquid with the velocity $\overset{\rightarrow}{\text{v}}$, and transmits (on
%the average) a momentum $mV$ to the solid surface.}}{\Qlb{fig2}}%
%{fig2.eps}{\special{ language "Scientific Word";  type "GRAPHIC";
%maintain-aspect-ratio TRUE;  display "USEDEF";  valid_file "F";
%width 2.6109in;  height 1.6985in;  depth 0pt;  original-width 4.3059in;
%original-height 2.7847in;  cropleft "0";  croptop "1";  cropright "1";
%cropbottom "0";  filename 'fig2.eps';file-properties "XNPEU";}}}%
%BeginExpansion
%EndExpansion

This leads to a shear stress:%

\begin{equation}
\sigma=mV_{s}\frac{\rho}{m}\overset{-}{\text{v}_{z}}=\rho V_{s}\overset
{-}{\text{v}_{z}} \label{eq4}%
\end{equation}

with:%

\begin{equation}
\overset{-}{\text{v}_{z}}=\int_{0}^{\infty}\frac{1}{\sqrt{2\pi\,}\text{v}%
_{th}}\text{v}_{z}e^{-\text{v}_{z}^{2}/2\text{v}_{th}^{2}}d\text{v}%
_{z}=\text{v}_{th}/\sqrt{2\pi} \label{eq5}%
\end{equation}

where v$_{th}^{2}=kT/m.$

Eq. (\ref{eq4}) gives us a precise value of the friction coefficient $k$ in
eq. (\ref{eq2}), and from it, we arrive at a Navier length:%

\begin{equation}
b=-h+\frac{\eta}{\rho\overset{-}{\text{v}_{x}}}\cong\frac{\eta}{\rho
\text{v}_{z}}\qquad(h<\ell) \label{eq6}%
\end{equation}

(The $-h$ term is du to the distance between liquid surface and solid surface;
it is completely negligible in practice).

If we choose typical values, $\rho=1$g/liter, \ \ v$_{th}=300$m/sec, and
$\eta=10^{-2}$poises, we find $b=$7 microns.\ Thus, a gas film can indeed give
a very large slip length. Our calculation assumed complete thermalisation at
each particle/boundary collision.\ If we had some non zero reflectance
(especially on the solid surface), this would increase $b$ even more.

\section{Conclusions}

When gas films with thickness $h$ in the interval $a<<h<<\ell$ are present in
a flow experiment, they may indeed generate a strong slippage, independent of
the film thickness.\ But the process which could generate such films remains
obscure.\ If, for some reason, the liquid entering the channel was
supersaturated with a certain gas, a pressure drop in the channel could indeed
promote the release of gas bubbles. If the solid surface is not very wettable
($\theta>90%
%TCIMACRO{\UNICODE{0xb0}}%
%BeginExpansion
{{}^\circ}%
%EndExpansion
$), the bubbles should preferentially nucleate at the liquid/solid interface
(as they do in a glass of champaign).\ Then, we would have to postulate that
the shear flow transforms them into a flat film, provided that the shear
stress is strong enough to flatten them. Indeed, in the Zhu-Granick
experiments, there is a threshold stress, but it is much smaller than what
would result from a compromise between shear stress and surface tension for
small bubbles.

It is worth emphasizing that the amount of gas required to lubricate the solid
liquid contact is very small: if $D$ is the macroscopic width of the shear
flow, the weight fraction of gas required is:%

\begin{equation}
\psi=\frac{\rho}{\rho_{L}}\frac{h}{D} \label{eq7}%
\end{equation}

where $\rho_{L}$ is the density of the liquid. Typically $\psi\sim10^{-5}.$

On the whole, we cannot present a complete explanation of the anomalous Navier
lengths based on gas films. But the films may possibly show up in some
instances; e.g. if the liquid is purposely oversaturated with gas: then eq.
\ref{eq6} may become useful.

\bigskip

\bigskip

\textit{Acknowledgments}: I benefited from stimulating discussions with
S.\ Granick, L.\ L\'{e}ger, H.\ Hervet and F.\ Wyart-Brochard.

\end{document}